\documentclass[12pt,a4paper]{article}

 \usepackage[dvips]{graphics}
 \usepackage{graphicx}
 \usepackage{afterpage}
 \usepackage{enumerate}
\begin{document}

 \title{\bf The Fe I 1564.8~nm line  \\ and the distribution
of solar magnetic fields }

 \author{\bf V.A. Sheminova}
 \date{}

 \maketitle
 \thanks{}
\begin{center}
{Main Astronomical Observatory, National Academy of Sciences of
Ukraine
\\ Zabolotnoho 27, 03689 Kyiv, Ukraine\\ E-mail: shem@mao.kiev.ua}
\end{center}

 \begin{abstract}
The distribution of magnetic field strength at different levels in the quiet solar
photosphere was obtained on the basis of the 2D MHD simulation of
magnetogranulation and the synthesis of the  V profiles of the Fe I
$\lambda$~1564.8~nm line. The shape of the distribution and the position of its
maximum vary essentially with depth. The distribution maximum lies, on the average,
at 25~mT, but it is found near 35~mT when the spatial averaging of profiles (about
$0.5^{\prime\prime}$) is taken into account. The difference between the
distributions is due to the errors with which the field strength is determined from
V profile splitting. Our analysis reveals that the use of the Fe~I $\lambda$~1564.8
nm line in this method is the most efficient and reliable means of measuring fields
above 50~mT, when this line is under the strong-splitting conditions. Under the
weak-splitting conditions the measured field strengths are about 20~mT, while under
intermediate conditions they are overestimated by 2--4~mT. The field strength
distribution obtained with the $\lambda$~1564.8~nm line in the fields above 50~mT
can serve as a standard for testing other techniques and lines. The analysis of the
synthesized Fe I  $\lambda$~630.2~nm line profiles supports the conclusion that
this line is less suitable for studying field strength distributions because of its
weak magnetic sensitivity to fields below 120~mT. In addition, the separation of
the $\sigma$-components of this line highly depends on the magnetic vector
inclination. The separation of  V peaks increases with vector inclination, and this
results in a considerable overestimation (by about 20--30~mT) of the strength of
weak inclined fields determined by the methods which ignore the Q and U profiles.
The derived magnetic field distributions as well as the distributions of asymmetry
parameters and  V profile zero-crossings are in good agreement with IR polarimetry
data. They convincingly confirm the assumption that the structure and strength of
photospheric magnetic fields of mixed polarity have continuous spectra down to
scales considerably smaller than the resolution threshold. They also suggest that
the structure and the scales of magnetic fields are closely related to the
granulation structure.
\end{abstract}

\section{Introduction}
     \label{S-Introduction}

In our previous studies of the solar magnetogranulation based on 2D MHD simulations
by Gadun \cite{Gadun00, Gadun01} we have touched upon the very important problem of
the distribution of magnetic fields in quiet photospheric regions. As the interest
in the weak magnetic fields of the quiet Sun, their nature and structure, and in
the interpretation of their observations continues to grow \cite{Gross96, Keller94,
Lin99}, we return to the analysis of the numerical simulation and the synthesis of
the Stokes profiles of the infrared line $\lambda$~1564.8~nm in the context of weak
photospheric magnetic fields.

Observations of magnetic fields in the quiet regions on the Sun evidence very weak
magnetic fluxes ($10^{-7}$--$10^{-8}$~Wb). The fine field structure cannot be
resolved even with the highest spatial resolution, a very high signal-to-noise
ratio, and at good seeing. It is also difficult to calibrate correctly the
measurements, i.e., to determine the true field strength. For the time being the
only possibility to study in detail the small-scale structure of photospheric
fields is the direct numerical simulation of magnetoconvection. Our high-resolution
studies with a spatial step of 35 km \cite{Gadun99, Shem99} made with the use of 2D
MHD magnetogranulation models \cite{Gadun00} demonstrated that these models are
helpful in understanding various properties of photospheric magnetic fields and
their interaction with convective motions on granular scales. These models, when
used to study the magnetic field distribution in granules and intergranular lanes,
will hopefully provide insight into various problems related to the structure and
nature of magnetic fields in the photospheres of the Sun and stars.

Among numerous observation data on the magnetic fields of the quiet Sun, of special
interest are the observations of IR lines which turned out to be very helpful in
measuring small-scale fields below 0.1 T. The results reveal a complex field
structure in quiet regions with low flux densities. The magnetic fields of the
quiet Sun seem to be strongly mixed in space -- weak, intermediate, and strong
fluctuations as well as thin fluxtubes alternate with one another. Theoretical
magnetoconvection simulation \cite{Emonet} also suggests that magnetic fields in
quiet regions are structurized from the greatest to the smallest ones. Their sizes
can be much smaller than the present-day resolution threshold. In addition, the
theory as well as observations suggest that the magnetic fields are a mixture of
fields of different polarities. Measurements of magnetic fluxes on the quiet Sun
outside the supergranulation network \cite{Lites02} indicate that the disbalance of
fluxes of different polarities there is smaller than in the network. Recent
advances in the observations of weak fields raised the questions as to whether the
weak fields are the remnants of a strong magnetic flux circulating at all times due
to convection or they are generated by the local dynamo mechanism, whether their
dimension and strength can vary continuously down to very small values, and whether
the data on the distribution of weak fields are sufficiently reliable. The
distributions derived from the observations of Stokes profiles of spectral lines in
the visible range and in the infrared turn out to be essentially different. The
maximum of the relative distribution obtained from the $\lambda$~1564.8~nm line
\cite{Collados01, Khom02, Lin95, Lin99} indicates that the majority of fields have
strength much lower than 100~mT, while the observations in the visible spectrum
give a maximum of the strength distribution at about 100~mT \cite{Gross96, Lites02,
Sanchez00, Socas02} and the fields below and above 100~mT fill only a small part of
the photosphere volume (about 1~percent) in quiet regions. According to the IR
observations, this part is slightly greater, but it is still quite small as
compared to the part occupied by very weak (0.4--4~mT) turbulent fields which cover
the whole surface of the quiet Sun (they were recently detected through the use of
the Hanle effect \cite{Stenflo98}). All these results obtained by various
techniques can be reconciled when we assume complex topology of magnetic fields in
the photosphere, where weak and strong different-scaled magnetic fields are mixed
and entangled. At the same time it is not clear why the field strength distribution
maximum in quiet photospheric regions points at kilogauss fields when lines in the
visible spectrum are observed and subkilogauss fields when IR lines are measured.
The distinction can be caused by the specificity of magnetic field measurements in
the visible and IR lines, which strongly differ in their parameters.

In this study based on a time sequence of 2D MHD magnetogranulation models
\cite{Gadun00, Gadun01}, we obtained a distribution of magnetic fields with a high
numerical resolution in the solar photosphere outside active regions. We
established a possible cause of the differences in the field distributions derived
with the use of synthesized visible and IR lines.

\section{Magnetic field strength distribution}

We took a 30-min sequence of 2D MHD models described in detail in \cite{Gadun00,
Gadun99, Gadun01}. The sequence contains 56 two-dimensional models with 112 columns
in each.  We extracted the field strength $B$ at several
photospheric levels ($\log \tau_5=0,$ -0.5, -1, -1.5, -2, -3) 
in every model column   and  built the field
strength distribution based on the simulation data. Figure 1 shows the strength
distributions at these levels (thick curves).

At the next step we synthesized the Stokes profiles of the $\lambda$~1564.8~nm line
for every column in every model by integrating the Unno-Rachkovskii equations for
the polarized radiation transfer. Thus we obtained the 6272 profiles  for various
statistical investigations of relative distributions of atmospheric parameters
derived from the Stokes profiles of the $\lambda$~1564.8~nm line.

Note that the choice of the IR $\lambda$~1564.8~nm line for our investigations  was
not accidental. The polarization properties of this line were discussed in detail
in \cite{Sanchez00}. The authors of \cite{Sanchez00} assume that the observations
of quiet regions on the Sun with magnetic elements with strengths between 50~mT and
150~mT will give  the field distribution  with predominantly 50~mT  when
$\lambda$~1564.8~nm is used and 100~mT when the visible line Fe I
$\lambda$~630.2~nm is used.  It is not difficult for us to verify these assumption,
since we have a nonhomogeneous photosphere model with known field distribution at
various levels, on the one hand, and the Stokes profiles calculated with the same
models which can be used to find the field strength distribution, on the other
hand. The observations of IR line profiles are often reduced with the use of a
simple method in which the field strength is determined from the distance between
the positive and negative maxima (peaks) of V profiles. We demonstrated in
\cite{Shem99} that this direct method gives the most accurate field strengths when
applied to the $\lambda$~1564.8~nm line, and so we used it in this case. We also
made similar calculations for the visible $\lambda$~630.2~nm line to be able to
compare the results for two different lines. We omitted from consideration the
abnormally shaped V profiles which have two or more zero-crossings and can thus
introduce additional errors in the distributions. The total numbers of analyzed
profiles were 4577 for $\lambda$~1564.8~nm and 3897 for $\lambda$~630.2~nm. Both
field distributions thus obtained are plotted in Fig.~1. The distribution obtained
with the visible line demonstrates complete disagreement with the true
distribution. The distribution calculated with the IR line coincides best with the
MHD models at levels $\log \tau_5=$ -0.5 and -1 for fields above 50~mT.

To specify the depth in the photosphere to which the field strength found from the
$\lambda$~1564.8~nm line should be referred, we calculated the effective depths of
V-peak formation. We used the contribution functions \cite{Shem92} to find the mean
level in the nonhomogeneous photosphere where the effective absorption in the
profiles of this line occurs. We calculated the profiles for two typical areas in
the model -- one area covers the periphery and a part of a strong fluxtube and the
other covers the center of a granule with predominantly horizontal fields. We
obtained virtually all types of profiles met in granules and intergranular lanes.
Figure 2 shows these I and V profiles together with the profiles of the effective
depth of line formation. This effective depth substantially varies from one model
column to another because of steep gradients of thermodynamic parameters along the
line of sight and the magnetic field and velocity field gradients. It is difficult,
therefore, to determine exactly the mean depth of line formation to which the field
strength distribution refers. Nevertheless, we may conclude from Fig.~2 that the
effective layer of the formation of V profile peaks lies at the level $\log
\tau_5=$~-0.5 for $\lambda$~1564.8~nm and at $\log \tau_5=$ -1 for
$\lambda$~630.2~nm.

\begin{figure}
   \centering
   \includegraphics[width=16. cm]{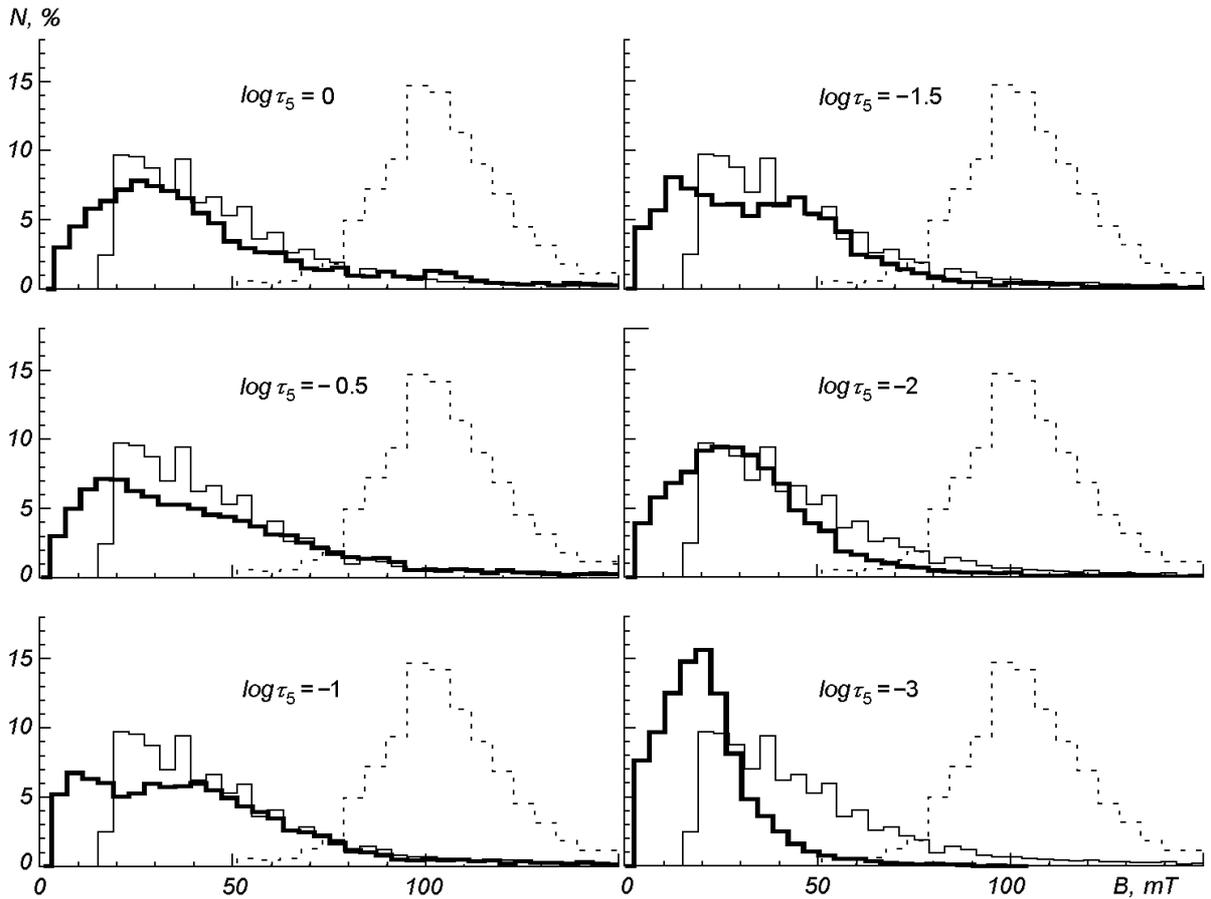}
   \caption[]{Magnetic field strength histograms
(thick line) plotted from the data obtained directly from the time sequence of 2D MHD
magnetogranulation models at various levels $\log\tau_5$ indicated on the plots. Thin and dotted
lines show histograms of magnetic field strength  obtained from the distances between 
the V peaks of the lines
$\lambda 1564.8$~nm and $\lambda 630.2$~nm, respectively.  }
      \label{Fig1}
\end{figure}

So, the field distributions we obtained from the MHD magnetogranulation simulation
and from the synthesis of the $\lambda$~1564.8~nm and $\lambda$~630.2~nm lines
allow us to draw the following conclusions.

1) In areas outside  active regions the majority of photospheric magnetic fields
are weaker than 50~mT and the field distribution maximum lies near 25~mT, on the
average.

\begin{figure}
   \centering
   \includegraphics[width=16.5 cm]{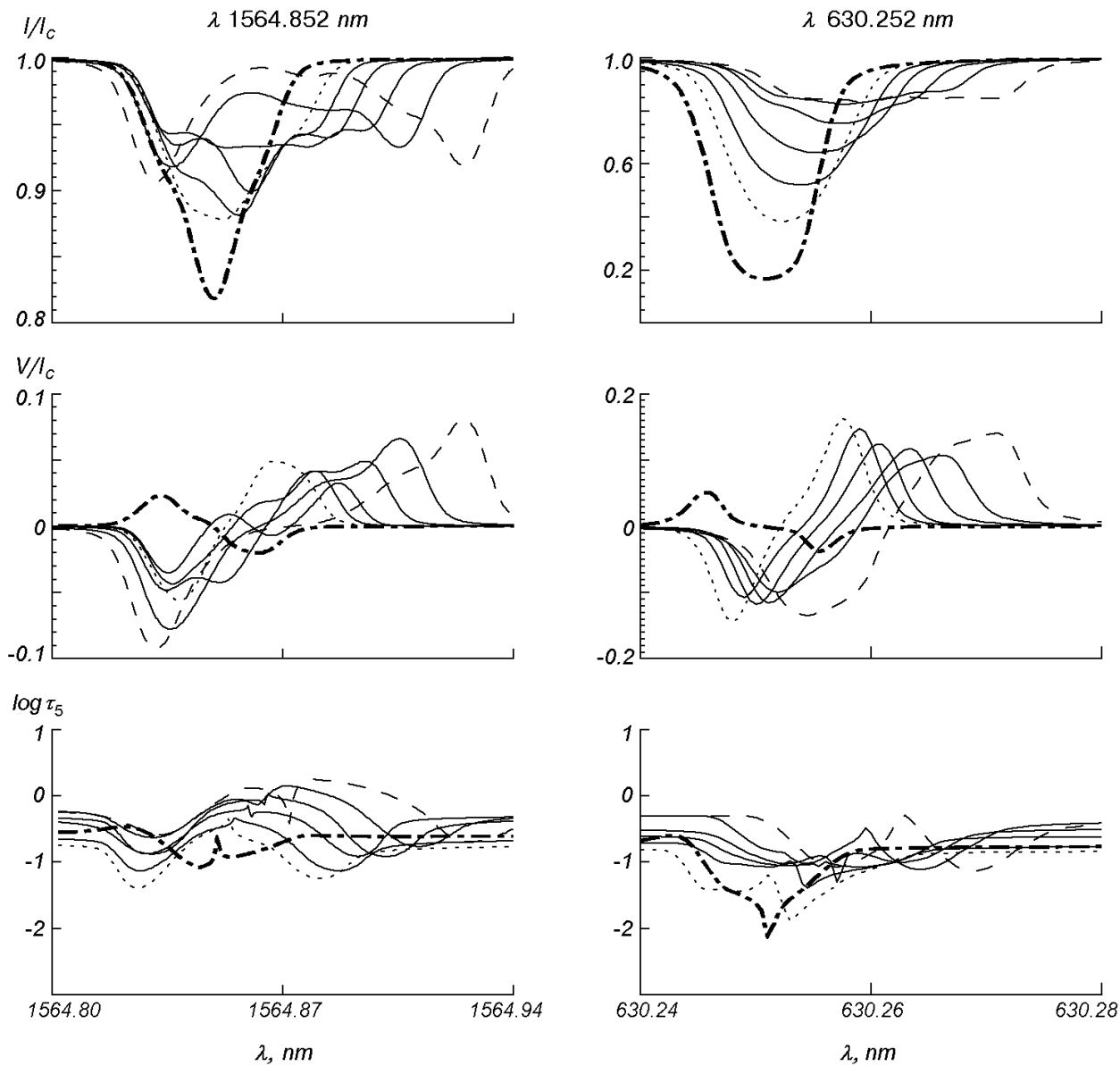}
   \caption[]{The Stokes I profiles (top panels) and the  Stokes V profiles (middle panels)  and
the effective depths of V profile formation (lower panels) for lines
$\lambda 1564.8$~nm and $\lambda 630.2$~nm calculated at a  periphery of the fluxtube
(solid line), its centre (dashed line),  its edge (dotted
line), and a granule centre  (thick dash-and-dot line).
 }
      \label{Fig2}
\end{figure}

2) The relative distribution of magnetic field strength changes noticeably at
different levels in the photosphere -- a redistribution of fields in height takes
place. The number of strong fields ($>100$~mT) decreases, and the distribution peak
changes its position. At the photosphere base ($\log \tau_5=0$) the principal
distribution maximum is indicative of the predominance of fields with strengths of
about 25~mT. There is also a smaller peak near 100~mT and an even smaller one near
70~mT. At the level $\log \tau_5=$~-0.5 the fields become weaker, and the
distribution has only one peak near 20~mT. At the levels $\log \tau_5=$~-1 and -1.5
the principal peak shifts toward weaker fields (about 10~mT), and a new peak of
about the same height appears near 45~mT. The doubling of the principal peak is the
manifestation of the canopy effect (strong fluxtubes expand with growing height).
As a result, the contribution of strong fields of fluxtubes in the total number of
fields increases. In still higher layers the field in fluxtubes expands even more
and becomes weaker as well, and the distribution of weak fields becomes nearly the
same as at the photosphere base. So, the MHD simulation results substantiate the
view that magnetic fields become weaker with growing height in the photosphere and
that the fields are subjected to redistribution.

3) Photospheric magnetic fields outside the active regions are a mixture of weak,
moderate, and strong fields. Their strength can vary continuously from the lowest
values in inclined fields which fill the entire granule region to the greatest
values in the fields compressed into thin vertical fluxtubes which are found in
intergranular lanes.

4) A comparison between the true distribution (obtained from MHD models) and the
calculated distribution (obtained from profile synthesis) clearly shows that the
method in which field strength is determined directly from the distance between the
V profile peaks of the $\lambda$~1564.8~nm line allows fields of 17~mT and greater
to be measured. The number of measured fields is overestimated in the interval
20--50~mT, but for fields above 50~mT it agrees satisfactorily with actual numbers.
It should be stressed that the number of fields with strengths of about 150~mT is
estimated quite reliably by this method.

5) The field strength distribution derived from the $\lambda$~630.2~nm line
suggests that this line is unsuitable for the given method of field strength
measurements. The distribution maximum corresponds to fields with $B
\approx100$~mT, which is at variance with the actual distribution.

We tried to find out why the distribution based on the $\lambda$~1564.8~nm line
profiles differs significantly from the actual distributions for fields below 50~mT
and why the $\lambda$~630.2~nm line is not suitable for such studies at all. With
this in mind we analyzed in detail the cause of the variation in the V profile peak
separation in nonhomogeneous models.

\section{The reliability of the field strength distribution \\
obtained from the 1564.8~nm and  630.2~nm lines}

Below we give some quantitative estimates for the accuracy of the field strength
determinations based on the $\lambda$~1564.8~nm and $\lambda$~630.2~nm lines.
Recall that the sensitivity of a line to a given magnetic field depends on the
width ratio $\Delta \lambda_B{/} \Delta \lambda_D$, where $ \Delta \lambda_B$
 is the Zeeman splitting and $\Delta \lambda_D$ is the Doppler line width in
the absence of magnetic fields. These widths depend on wavelength, temperature,
line saturation, nonstationary velocities and their gradients. The Zeeman width
also depends on the Land\'{e}  factor as well as on the magnitudes and gradients of
field strength and magnetic vector inclination. Hence it follows that the quantity
 $\Delta \lambda_B{/} \Delta \lambda_D$ for a specific magnetic field
and a specific velocity field is a function of temperature and it grows
approximately linearly with wavelength $\lambda$~and Land\'{e} factor $ g_{\rm
eff}$. The product $\lambda g_{\rm eff}$ is often taken as a measure of the
sensitivity of spectral lines to magnetic field. In the atmospheric regions with
widely different magnetic properties the trustworthiness of the field strength
measured with the use of Zeeman splitting in a specific line depends on the
magnetic sensitivity of the line (e.i., from $\lambda g_{\rm eff}$) as well as on
the magnetic field parameters, in other words, it depends on the splitting
conditions for the given line. The splitting conditions which are illustrated in
\cite{Rbin92, Solanki93, Solanki92, Stenflo87} are commonly divided into three
groups. Soft conditions correspond to $\Delta \lambda_B < \Delta \lambda_D$,
intermediate conditions to $\Delta \lambda_B \approx \Delta \lambda_D$, and hard
ones to $\Delta \lambda_B > \Delta \lambda_D$. Special test calculations were made
to establish these conditions for the two above lines and for our model atmosphere.
We calculated the line profiles for one of the MHD model columns, having replaced
the depth-dependent model field strengths by constant ones which varied from their
smallest value to the greatest one. We also assumed that the fields were
longitudinal. We calculated the field strength $B_{br}$ from the measured distances
between the blue and the red V profile peaks and compared it to the actual field
strengths (solid curves in Fig.~3). One can see that the conditions of strong
splitting occur at $B > 30$~mT for the $\lambda$~1564.8~nm line and at $B > 150$~mT
for the $\lambda$~630.2~nm line. The splitting is weak for all fields with $B <
17$~mT for $\lambda$~1564.8~nm and $B < 60$~mT for $\lambda$~630.2~nm. Under soft
conditions, the measured field strength is determined by the Doppler width only and
does not depend on any B variations -- the threshold field strength is close to
20~mT for $\lambda$~1564.8 nm and 90~mT for $\lambda$~630.2~nm. In the case of
intermediate conditions, the overestimate of field strengths is greater for weaker
fields. We also examined the influence of the gradients of magnetic field,
temperature, vertical velocity, and inclination on the conditions which set in one
case or another. It turned out that the action of field gradient on profile shape
is greater than on the distance between $|V|$ maxima. This distance is also only
slightly affected by temperature and velocity gradients, but the field inclination
exerts the greatest effect on the separation of V profile peaks. As seen from
Fig.~3, an increase of field inclination extends the limits of the intermediate
condition interval towards stronger fields, and the hard conditions set in at
$B\approx 40$~mT for $\gamma=30^{\circ}$ and $B \approx 50$~mT for
$\gamma=75^{\circ}$ in the case of the $\lambda$~1564.8~nm line. This effect is
even stronger for $\lambda$~630.2~nm. As a result the field strength is
overestimated still further, and the deviation of the measured field strengths from
the actual ones is greater in more horizontal and weaker fields.

\begin{figure}
   \centering
   \includegraphics[width=15.5 cm]{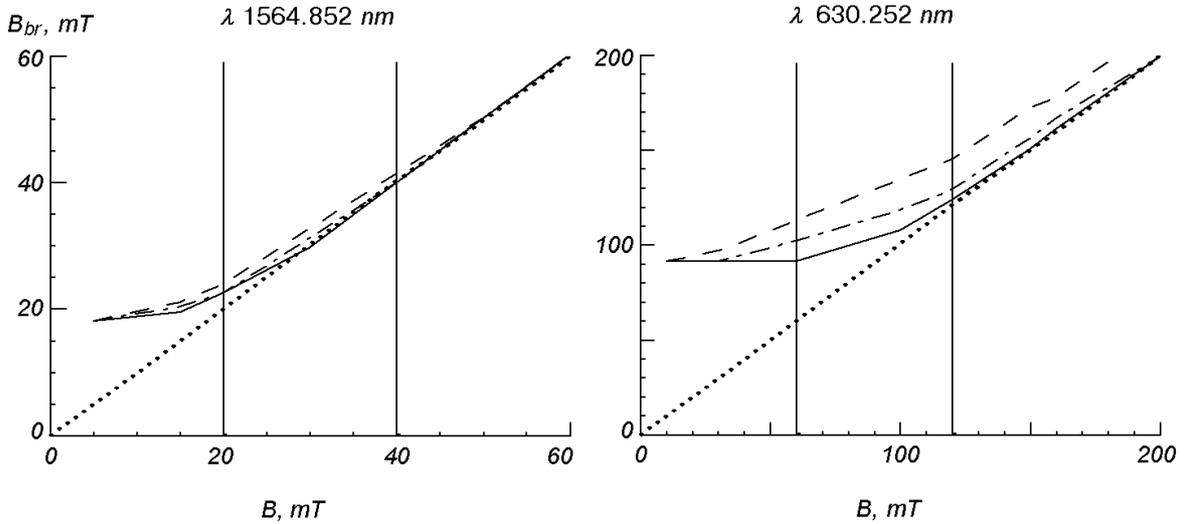}
   \caption[]{The  field strengths $B_{br}$ derived from the distances
between the V peaks of the synthesized profiles $\lambda 1564.8$~nm and $\lambda
630.2$~nm  vs. the true (model) field strengths $B$ for  magnetic vector
inclination of $0^\circ$ (solid line),  $30^\circ$ (dash-and-dot line), 
and $75^\circ$ (dashed line). Dotted line is line of equal field strengths. 
Vertical lines indicate the field strengths at which 
$\Delta \lambda_B {/} \Delta \lambda_D$ is equal to 1
(left line) and 2 (right line).
 }
      \label{Fig3}
\end{figure}

So, the direct method for the determination of $B$ from the distance between the
$|$V$|$ profile maxima works better in the case of strong and longitudinal fields,
and its accuracy deteriorates as the magnetic vector deviates from the vertical.
For lines in the visible spectral range (like $\lambda$~630.2~nm) the measured
field strength can be overestimated by 20--40~mT for strongly inclined fields. The
line $\lambda$~630.2~nm can give trustworthy results only for strong-splitting
conditions (fields above 150~mT) with inclinations less than $30^{\circ}$. In the
case of weak-splitting conditions all measured fields will have a strength of about
100~mT. At the same time the effect of field inclination on measured strengths is
insignificant for the $\lambda$~1564.8~nm line. The greatest overestimate which is
possible under intermediate-splitting conditions is about 2--4~mT for strongly
inclined fields. The field inclination effect for $\lambda$~1564.8~nm is 10 times
weaker than for $\lambda$~630.2~nm.

The accuracy of field strength measurements reflects directly on the field
distributions, especially on those obtained for quiet regions, where inclined and
weak fields dominate. In the case of soft conditions, when the magnetic sensitivity
of lines is weak, weaker fields are measured as stronger ones (especially in the
$\lambda$~630.2~nm line). As a result the field distribution is considerably
distorted in the region below 150~mT (for $\lambda$~630.2~nm), while in the
distribution derived with the use of the $\lambda$~1564.8~nm line the region of
weak fields between 20 and 40~mT is substantially distorted and the fields below 17
mT are absent at all. It is precisely these results that we obtained (see Fig.~1).

Hence it follows that in the case of complete resolution (filling factor
$\alpha=1$) the most trustworthy field distribution is that which was found from
the distance between the V peaks of $\lambda$~1564.8~nm only under strong-field
conditions, i.e., for strengths above 50~mT. The number of fields measured in the
range -20--50~mT is overestimated: first, weak fields below 17~mT are measured as
20-mT fields and, second, the presence of inclined fields results in an
overestimation by 2--4~mT. So, the main maximum of the field distribution found
from the $\lambda$~1564.8~nm line for quiet regions on the Sun is expected to be
slightly shifted toward stronger fields with respect to the actual distribution,
and it should also be higher at the expense of the fields from the weak-field
conditions. It is most likely to be located at 30--40~mT. The field distribution
beyond 50~mT up to the strongest kilogauss fields is not distorted.

In the cases when the spatial resolution is not high (filling factor $\alpha < 1$),
there is an additional profile broadening caused by the horizontal averaging of the
profiles. This broadening also affects the shape of the distribution of measured
fields, as judged from Fig.~4, which displays some distributions obtained from the
$\lambda$~1564.8~nm line (thick curve), and directly from MHD models (dotted line).
The greater the spatial averaging scale, the greater is the shift of the
distribution maximum toward stronger fields (30--40~mT); in addition, another
maximum appears at 50--60~mT, the number of fields above 60~mT noticeably
decreases, and the distribution shows a steeper decline toward stronger fields. In
Fig.~4 we also give for comparison the observation data from \cite{Khom02} acquired
with a resolution of 0.5--1$^{\prime\prime}$. The best agreement can be seen in the
second panel, where the profile averaging seems to be close to the angular
resolution of observations. Satisfactory agreement between our calculations and
observations from \cite{Khom02} argues for the reliability of our calculations and
distributions of weak fields on the quiet Sun.

As to the $\lambda$~630.2~nm line, the distances between its peaks give a
completely false distribution. This line can be used only to measure longitudinal
fields above 150~mT. All fields below 100~mT are fixed as fields of 90--120~mT, and
the resulting distribution has its maximum at $B=100$~mT. In actual practice this
line is not used in the method discussed here. It is often used, as a rule, in the
inverse codes, when a model atmosphere and a magnetic vector are found by comparing
the observed and calculated V profiles. Our analysis shows that it is practically
impossible to recognize the separation of the V peaks in weak profiles without
additional analysis of the Q, U profiles at small field strengths and large
magnetic vector inclinations. The inverse methods can give stronger longitudinal
fields instead of weaker horizontal fields. It is not surprising, therefore, that
the distributions obtained by inverse methods with the $\lambda$~630.2~nm line have
their maxima, as a rule, at 50--100~mT. We believe that the inverse methods need to
be tested with the 2D MHD or 3D MHD models. Only then one may judge on the accuracy
of the inversion.

\section{Statistical properties of photospheric \\ magnetogranulation
from the  1564.8~nm data}

We consider the statistical relationships between the granule structure parameters
derived from the synthesized $\lambda$~1564.8~nm line profiles. Figures~4a and 4b
display the field strength $B_{br}$ histograms and $B_{br}$ averaged over equal
intervals as a function of the line-of-sight velocity  $V_z$ derived from the V
profile zero-crossing shift.  Note that negative velocities
correspond to downward motion.  The field increases with the downflow velocity, and
this suggests that the magnetic field becomes stronger in the intergranular lanes,
where downflows are concentrated. The same is illustrated by the relation in
Fig.~4c which points to an increase of field strength in darker photospheric
regions where the contrast $I_c/{<}I_c{>}$ is less than unity. The field strength
changes from $40\pm25$~mT to $90\pm40$~mT, on the average, when going from granules
to intergranular lanes. Similar relations derived from observations \cite{Khom02}
give 45~mT for granules and 69--80~mT for the lanes.

The radial velocity $V_z$ histograms (Fig.~5a) give a mean velocity of 0.37~km/s,
which suggests the predominance of intense downflows. These downflows are
associated with intergranular lanes, as suggested by close correlation between mean
velocity and intensity contrast (Fig.~5c). A close relation between velocity and
inclination should be also noted (Fig.~5b). Fields with greater inclinations are
met mainly in the region of upflows, at sites with higher contrast, that is, at
bright centers of granules. The asymmetry of radial velocity histograms and the
correlation between velocity, inclination, and contrast bear out the fundamental
property of magnetoconvection -- the asymmetry of upward and downward flows of
matter with frozen-in magnetic fields. As a result the magnetic field structure and
the scale of magnetic field variations are closely related to the granulation
dimensions and structure.

\begin{figure}
   \centering
   \includegraphics[width=16.cm]{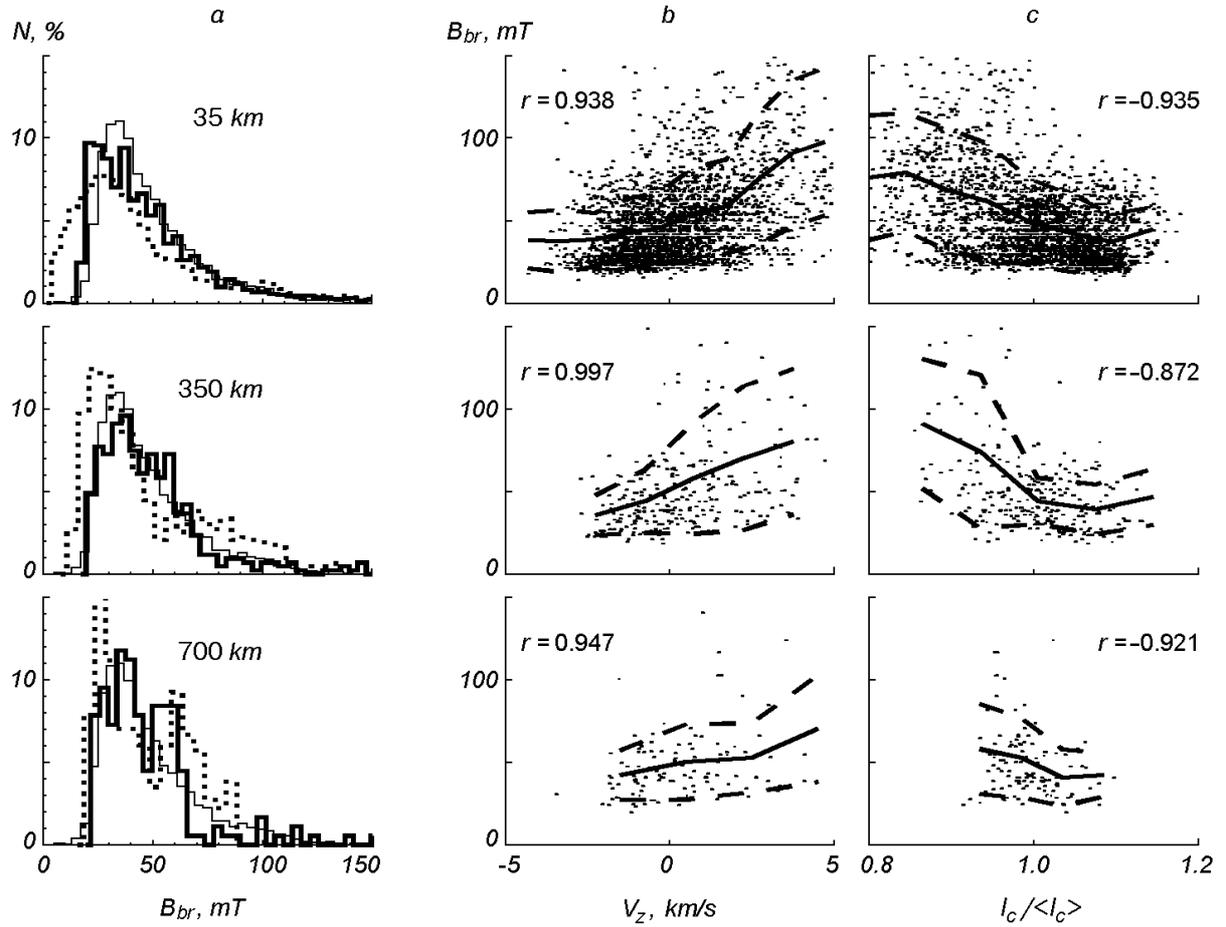}
   \caption[]{Magnetic field strength histograms (a) plotted  with the use
of the data from the synthesized $\lambda 1564.8$~nm line V profiles (thick line) and
MHD models (dotted line)  with different spatial averaging  (35, 350, 700 km). 
The date of observations  \cite{Khom02} with spatial resolution of 350--700~km idicate by thin line. Scatter plots  of the
field strength  vs. radial velocity (b) and  continuum intensity contrast
(c). Solid lines are correlation curves, dashed lines are rms deviations; $r$ is
the correlation coefficient. 

 }
      \label{Fig4}
\end{figure}


\begin{figure}
   \centering
   \includegraphics[width=16. cm]{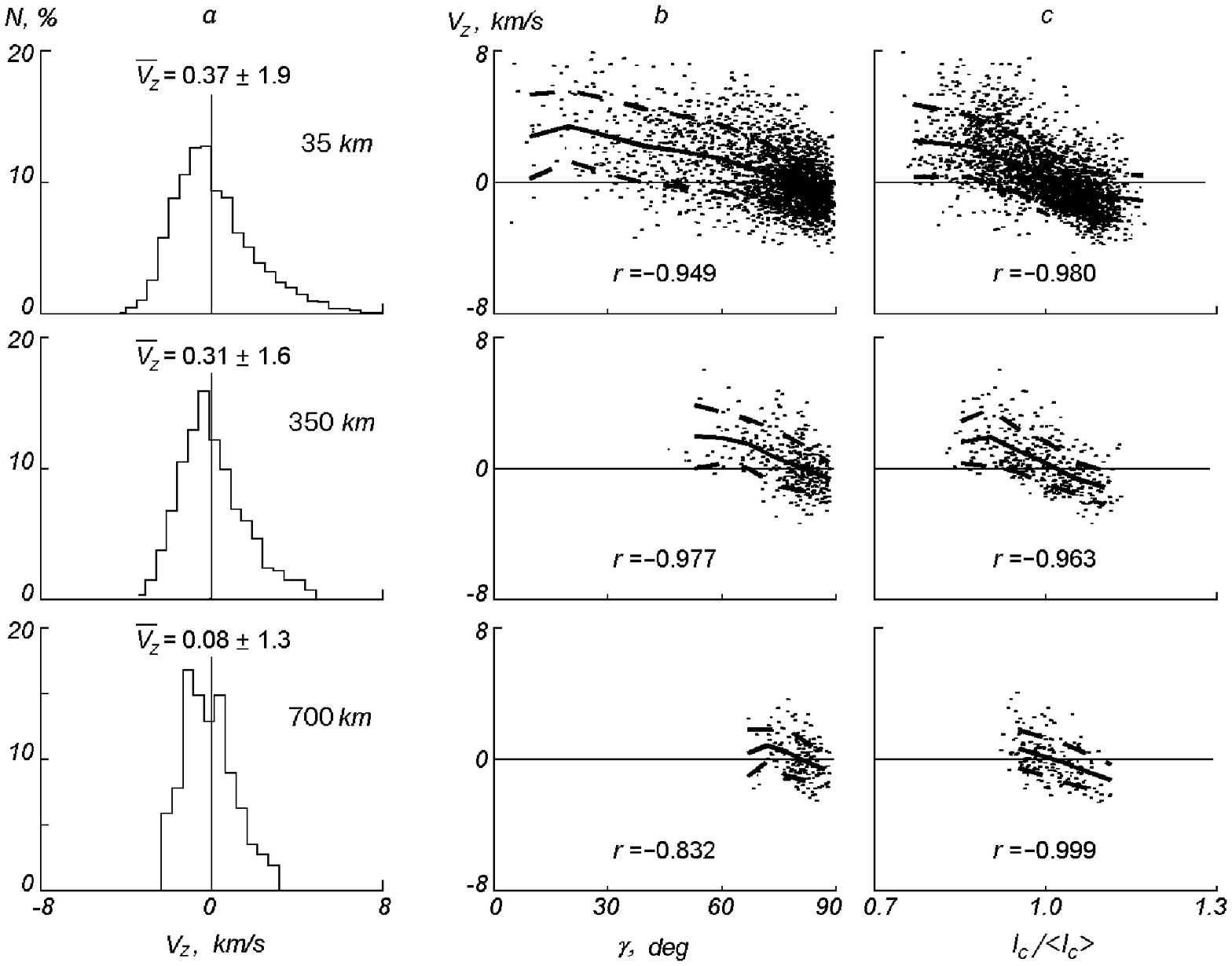}
   \caption[]{Histograms of radial velocities (a) derived from
    zero-crossing shifts of the synthesized $\lambda 1564.8$~nm line V
profiles. Scatter plots of the  radial velocities vs. magnetic vector
inclination $\gamma$ derived from the relation  $ \tan^ 2
\gamma= {(Q^2 + U^2 )}^ {1/2}{/} V^2$  (b),  and  continuum intensity contrast
$I_c/{<}I_c{>}$ (c). Symbol and curve coding are as  in Firure 4.
 }
      \label{Fig5}

\end{figure}

As all observation data contain the results of the analysis of the asymmetry of
observed Stokes profiles, we also give the corresponding data obtained from the
synthesis of the $\lambda$~1564.8~nm line V profiles in the form of histograms and
scattering plots as well as the relations between the corresponding quantities
averaged over equal intervals (Figs 6 and 7). To demonstrate the asymmetry, we took
the standard parameter $\delta a = (a_b - a_ r){/}(a_b + a_r$) which characterizes
the amplitude asymmetry between the blue and red peaks of V~profile. The
sample-averaged asymmetry is positive and very small (Fig.~6a), but it grows with
the scale of the spatial averaging of the profiles and approaches the observed
values \cite{Khom02}. The $\lambda$~1564.8~nm line asymmetry can change by $\pm40$
percent, much less than in the $\lambda$~630.2~nm line. This is evidence for a
lower sensitivity of the former line to temperature and velocity gradients and
magnetic vector variations. The area asymmetry $\delta A$ is measured in the same
way as $\delta a$, and the relation between these quantities is demonstrated in
Figs 6b,c. Figure 7 shows the correlation between the amplitude asymmetry $\delta
a$ and other parameters. The smaller the field strength, the greater is the scatter
of asymmetry parameter (Fig.~7a). The predominance of positive asymmetry tends to
increase as the field strength grows (Fig.~7b). There is no well-defined relation
between the asymmetry and radial velocity (Figs 7c,d), we can only notice that
there are more profiles with positive asymmetry in upflow regions (negative
velocities), while the number of profiles with negative asymmetry is slightly
larger in downflow regions. This weak relationship breaks down as the
spatial-averaging scale increases.

\begin{figure}
   \centering
   \includegraphics[width=16. cm]{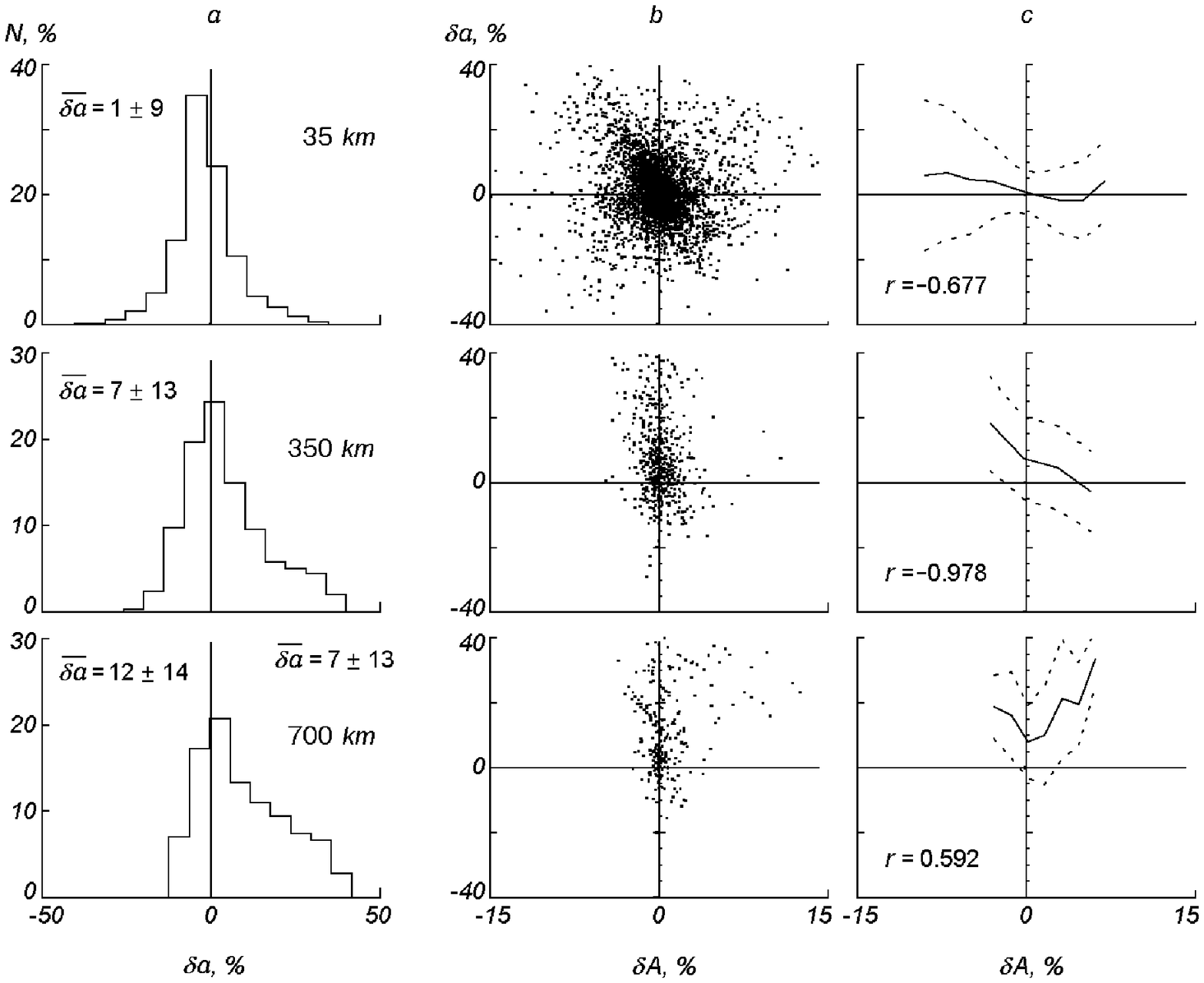}
   \caption[]{Histograms of the V profile amplitude asymmetry of
the synthesized $\lambda 1564.8$~nm lines (a);  scatter plots of the V profile amplitude
asymmetry vs. the V profile area asymmetry (b); solid lines are corresponding correlation
curves, dotted lines are rms deviations, and $r$ is the correlation coefficient
(c).
 }
      \label{Fig6}
\end{figure}

\begin{figure}
   \centering
   \includegraphics[width=16. cm]{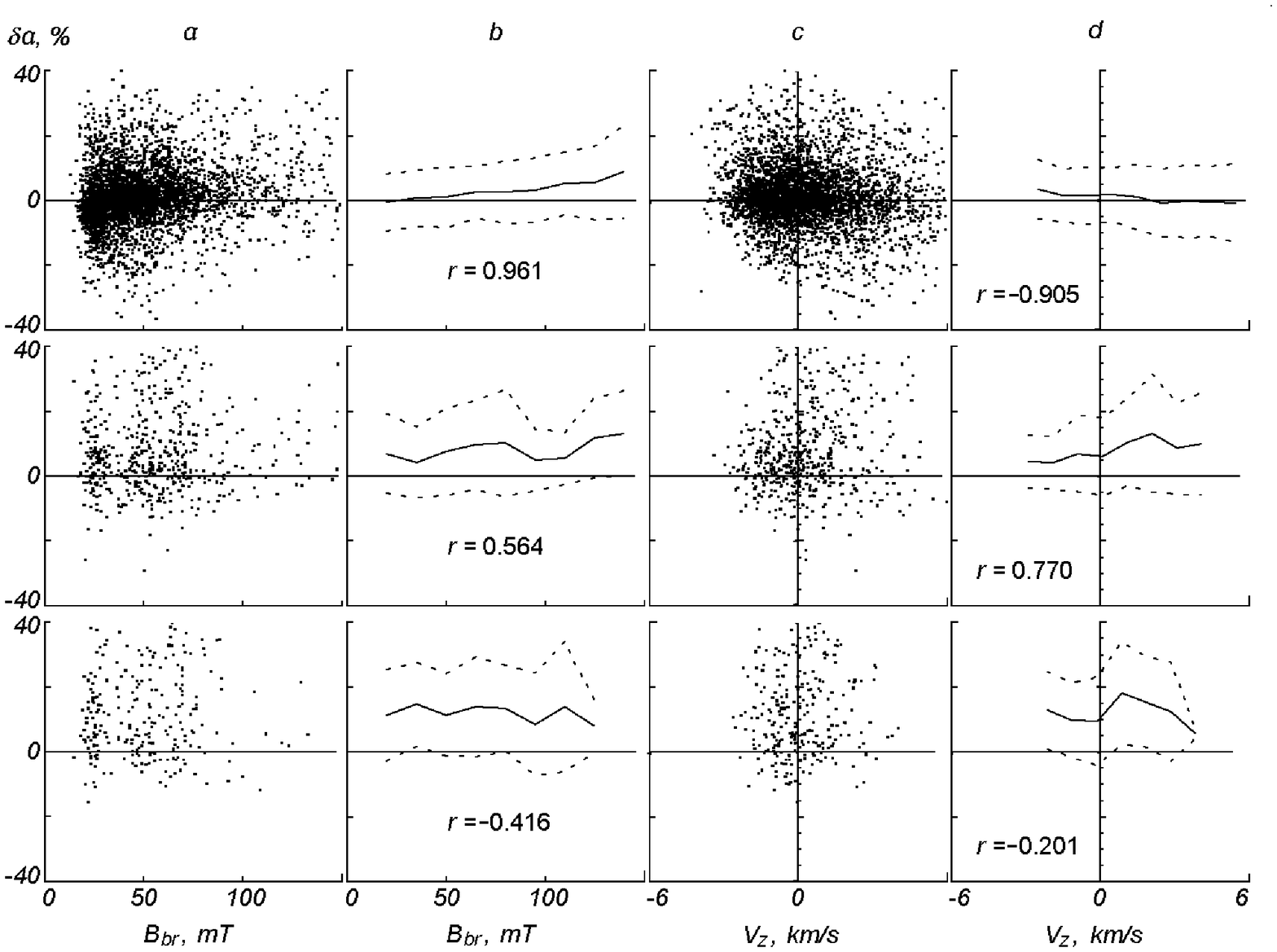}
   \caption[]{Scatter plots of the V profile amplitude
asymmetry of the synthesized $\lambda 1564.8$~nm lines vs. magnetic field strength (a),
radial velocity (c); corresponding correlation curves (b, d).  Symbol and curve
coding are as  in Firure 6.
 }
      \label{Fig7}
\end{figure}
%

Note that Figures 4--7 also display the statistical relations for the parameters
derived from the profiles averaged over 350-km and 700-km areas which correspond to
resolutions of $0.5^{\prime\prime}$ and $1^{\prime\prime}$. The number of profiles
drastically diminishes as the averaging scale grows, and the statistics results are
less accurate, although the effects of the horizontal averaging of profiles can be
still noticed in the figures. In some cases the results from the averaged profiles
are in better agreement with observations.

\section{Disbalance and distribution of magnetic flux}

Disbalance of positive and negative magnetic fluxes is an important aspect of
magnetic field distribution in the solar photosphere. It is defined as $\Delta F =
(F^{+} + F^{-}){/}(|F^{+}| + |F^{-}|)$ \cite{Lites02}, i.e., it is not affected by
unresolved fluxes, since the spatial resolution acts in the same way on both flux
components. We used this formula to calculate the magnetic flux disbalance in the
model region as a function of time. The magnetic flux was calculated every minute
for the 2D MHD models from  the whole 30-min 2D MHD model sequence as $F= \Sigma_{i
= 1}^{N} B_i(\log \tau_5 = 0)\Delta x^2$, where $i$ is the model column number,
$N=112$ is the number of columns in a 2D MHD model, and $\Delta x = 35$~km is the
column width. The field strengths $B_i$ at $\log \tau_5 = 0$ were taken from the
data of magnetoconvection simulation \cite{Gadun00}.

Figure 8 shows the $\Delta F$  variation for 30 min. The smallest
disbalance $\Delta F$ is 0.001 and the greatest one is 0.32. Time variations of
$\Delta F$ display oscillatory behavior.

Figure~9 (top panel)  shows the magnetic flux histogram derived from the whole
30-min MHD model sequence with $\Delta x = 35$~km.  The disbalance  $\Delta
F=-0.017$, the averaged field strength, or flux density, $\overline{B}= 0.202$~mT,
and the averaged unsigned field strength $\overline{|B|}=32$~mT. Figure~9   also
shows our results obtained with other spatial resolution $\Delta x = 350$~km
(middle panel) and $\Delta x = 700$~km (lower panel). The shape of flux
distribution  vary substantially, while the disbalance and mean field strength show
insignificant variations.

We also try to compare our results with the measurements data of the general
magnetic field (GMF) of the Sun \cite{Kotov77}, where the longitudinal field
component for the Sun as a star is given.  According to \cite{Kotov77}, the GMF
strength characterizes the overbalance of the magnetic field of one polarization
over the flux of the opposite polarization referred to a unit surface of the
visible disk.  Furthermore, the GMF is mainly controlled by the magnetic fluxes
from vast areas which are not related to active regions. The contribution of active
regions to the GMF is insignificant. The fields in such vast quiet areas were
called background fields. The GMF strength (or the background fields) is
0.1--0.2~mT. We obtained $\overline{B}=-0.2$~mT, i.e., the absolute value the same
as for the upper limit of observed field strengths. Although the agreement is quite
good, it could be worse for various reasons. The measured GMF critically depends on
magnetograph sensitivity and noise level. A twofold increase in the sensitivity
results in nearly the same increase in the measured magnetic flux. Besides, the use
of new calibration techniques in magnetic measurements also increases the
measurement result (by a factor of 2.4) because old techniques did not take into
consideration various factors: the saturation of lines in strong fields, when the V
signal is not proportional to magnetic flux any more, the weakening of lines in
regions with strong magnetic fields, and a partial compensation of the flux because
of the presence of opposite polarities in the resolved surface element. In general,
the measurement of the magnitude of the total solar flux with magnetographs with
commonly low spatial resolution is assumed to give underestimated fluxes.

The present measurements of the photospheric  magnetic fields with substantially
higher accuracy \cite{Lites02, Martin88, Wang95} found their complex structure. The
main components of these fields are network (N) and intranetwork (IN) fields. The N
fields have strengths of about 100~mT and mixed polarities, they are mainly met at
the corners of convective cells of supergranules and can form clusters. At solar
minimum periods the network covers the solar surface. According to current
measurements, the magnetic flux from network clusters is about $2\cdot
10^{10}$--$3\cdot 10^{11}$~Wb.

\begin{figure}
   \centering
   \includegraphics[width=10. cm]{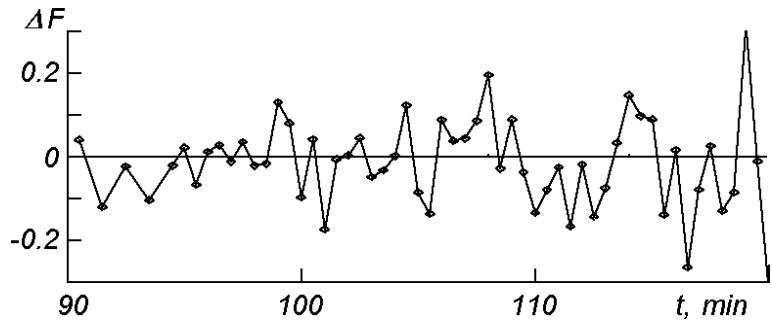}
   \caption[]{Magnetic flux disbalance calculated every minute
from  the  30-min 2D MHD model sequence as a function of simulation time.
 }
      \label{Fig8}
\end{figure}

\begin{figure}
   \centering
   \includegraphics[width=8. cm]{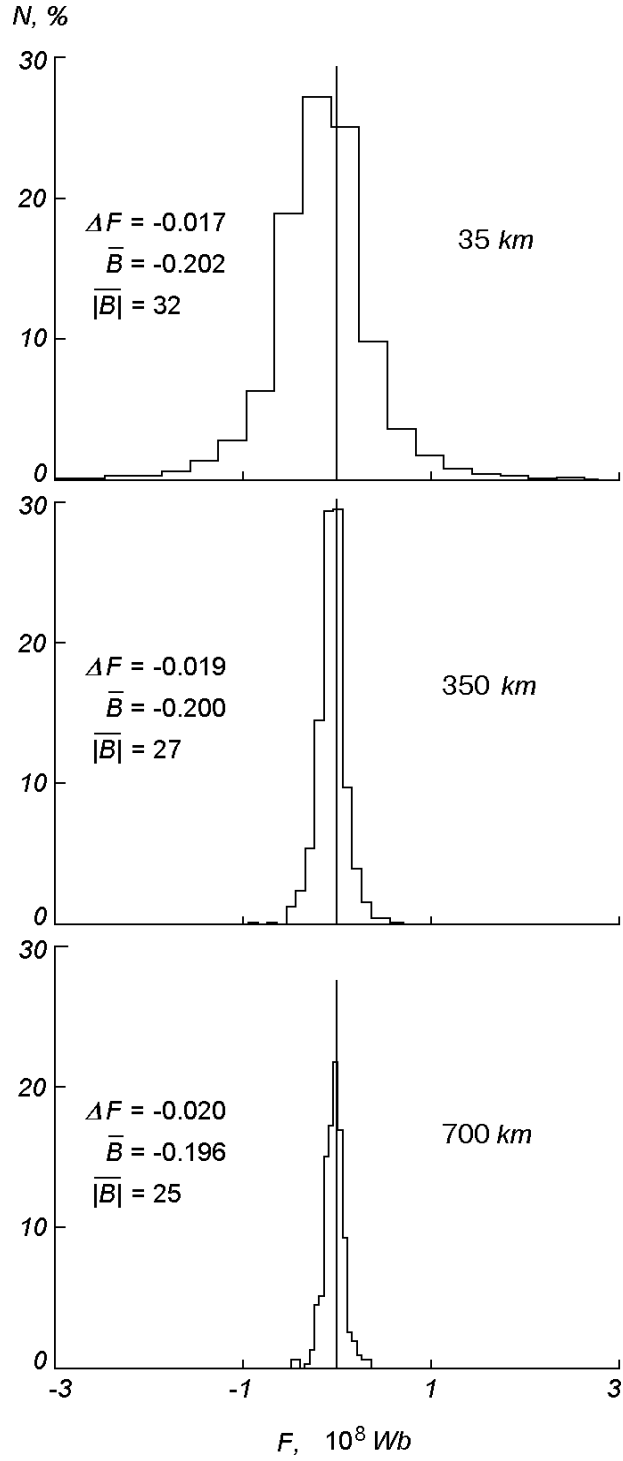}
   \caption[]{Magnetic flux distribution calculated with different spatial averaging
(35, 350, 700~km) for the 30-min 2D MHD  model sequence. Averaged field
strengths are given in~mT units.
 }
      \label{Fig9}
\end{figure}
%

With high sensitivity of polarimetric measurements and high spatial resolution at
good seeing, weak fluxes from IN fields can be measured separately from
the N fields. The IN fields are found inside the supergranulation network as big
and small fragments. They are weaker and more diffuse, their fluxes are a factor of
$10^2$ smaller than the fluxes from N fields. The mobility of IN fragments is
higher, they are more free to move from the center of a supergranule to its
boundaries. They can merge, annihilate one another, or interact with N fields.
Being weak and having mixed polarities, the IN fields cannot penetrate to the outer
chromosphere and corona. Observations of quiet regions with an spatial resolution
of about $0.5^{\prime\prime}$ \cite{Lites02, Wang95} revealed that the mean flux
density varied there from 0.3 to 4~mT. The IN fields contribute 0.165~mT to the
mean flux density, which corresponds to a total solar flux of $10^{15}$ Wb. Since
the lifetime of IN fragments is much less than one day, a flux of about
$10^{16}$~Wb emerges from the Sun and disappears in the form of IN fields in one
day. This flux is close to the total flux from the whole Sun observed at the
maximum of solar cycle 21. According to \cite{Wang95}, the disbalance of the IN
field flux is 0.08 and the mean field strength is 0.59~mT, while the IN disbalance
obtained in \cite{Lites02} varies from 0.03 to 0.48 depending on fragment observed,
and it is smaller approximately by a factor of three than the N disbalance. Our
magnetoconvection simulation gives  the mean field strength of 0.2~mT and
the flux disbalance of 0.02. It is less than the lower limit of the disbalance
observed with high spatial resolution. The difference seems to be caused, on
the one hand, by the fact that we use the results of  2D MHD simulation instead
3D MHD sinulation  and, on the other hand, by insufficient
accuracy of observations because of inadequate noise level and seeing conditions.

It should be noted that the extent of disbalance is also of importance in
identifying the sources of weak IN fields \cite{Lites02}. The IN fields differ from
the N fields not only by their properties but by their nature as well. As their
disbalance  is smaller, they are assumed \cite{Lites02} to originate from the local
dynamo and not from the global convection circulation appropriate to the network
fields. The authors of \cite{Stein02} ruled out the local surface dynamo as a
mechanism of the IN field generation. The energy of weak small-scale fields in the
surface layers grows due to field concentration rather than through the dynamo
action, and the concentration of fields, their stretching and twisting are caused
by convective motions. A 3D magnetoconvection simulation \cite{Stein02}
demonstrated that under the conditions of strong stratification and asymmetric
convective flows on the Sun a local short-period recirculation can arise near the
surface. Such is the mechanism which can constantly sustain weak magnetic fields
inside granules, mesogranules, and supergranules on the quiet Sun. Such a surface
recirculation was also obtained in the study \cite{Ploner01} based on the 2D
magnetoconvection simulation \cite{Gadun00}. The local surface short-period
recirculation is favorable to rapid mixing of the fields of opposite polarities on
small scales, thus diminishing the flux disbalance in the intranetwork fields.
Today, the question of the nature of the magnetic IN field is open.

\section{Conclusion}

We used 2D MHD magnetogranulation models  \cite{Gadun00} with relatively high
numerical resolution to investigate the magnetic field distribution in the solar
photosphere outside the active regions and obtained the following results.

1. At the photosphere base the magnetic fields are predominantly weak, their
strength is less than 50~mT. The strength distribution peak corresponds, on the
average, to 25-mT fields. The distribution tail is rather long, it extends to 150
mT. There is another small peak at 100--110~mT, it suggests that such strengths
dominate in kilogauss fields in  network fluxtubes. The photospheric field
distribution varies considerably with height, and this is evidence of a
redistribution of the fields due to their weakening, on the one hand, and the
expansion of fluxtubes, which increase the number of strong fields, on the other
hand. All these data suggest that photospheric magnetic fields are a mixture of
various fields, from the weakest ones to the strongest kilogauss fields of
fluxtubes. Their strength can vary almost continuously from the lowest values in
inclined fields in granules to the highest ones in thin vertical fluxtubes found in
intergranular lanes. Alternating flux polarities produce a polarity disbalance
about -0.02.

2. The direct measurement of magnetic field strength from the distance between the
maxima in the $\lambda$~1564.8~nm line profiles, providing a high spatial
resolution ($\leq 0.5^{\prime\prime}$), is a very efficient and reliable tool for
fields above 50~mT, but it fails for fields below 20~mT. All weak fields below 17
mT are measured as fields of 18--20~mT. In addition, the field strengths below 50
mT can be overestimated by 2--4~mT due to the effects of inclined fields which are
ignored in this method. So, the relative distribution of fields above 50~mT
acquired with the use of the $\lambda$~1564.8~nm line with sufficiently high
spatial resolution is close to the actual distribution and can serve as a standard
in testing other spectral lines with this method or in testing other methods. The
number of fields of 20--50~mT is always overestimated, and the existence of fields
below 20~mT in quiet solar regions remains unnoticed in these distributions. Even
with the complete resolution of magnetic fields this method applied to the
$\lambda$~1564.8~nm line, which is very sensitive to magnetic field, is unsuitable
for measuring very weak fields below 20~mT.

3. The inverse methods in which the strengths of weak magnetic fields are
determined from the Stokes V profiles of the visible lines like $\lambda$~630.2~nm
also seem to overestimate the field strengths as compared to the methods based on
the IR lines like $\lambda$~1564.8~nm. A possible reason is the weak magnetic
sensitivity of $\lambda$~630.2~nm to fields below 150~mT, on the one hand, and its
high sensitivity to magnetic vector inclination, on the other hand. It is difficult
to separate the field strength effect from the field inclination effect on the V
profile shape and the distance between profile peaks without making recourse to the
Q and U profiles. The field inclination effect for the $\lambda$~630.2~nm line is
nearly 10 times stronger than for the $\lambda$~1564.8~nm line. Therefore, the
field distributions found with the use of the $\lambda$~630.2~nm line for quiet
photospheric regions, where the fields are predominantly inclined, are less
reliable, especially for subkilogauss fields, compared to the distributions found
with the $\lambda$~1564.8~nm line. The high sensitivity of $\lambda$~630.2~nm to
magnetic vector inclination can be successfully used for the diagnostics of this
field parameter.

{\bf Acknowledgements.}We wish to thank S. Solanki and E. Khomenko for making
available the observation data on magnetic field distributions. This study was
carried out within the framework of an international research program and was made
possible in part by the INTAS Grant No. 0084.



\end{document}